\documentclass[%
 reprint,
 amsmath,amssymb,
 aps,
prab,
]{revtex4-2}

\usepackage{graphicx}
\usepackage{dcolumn}
\usepackage{bm}
\usepackage{hyperref}
\usepackage{amsmath}

\usepackage[authormarkup=none]{changes}

\usepackage{float}

\begin{document}

\preprint{APS/123-QED}

\title{Compact polarized X-ray source based on all-optical inverse Compton scattering}

\author{Yue Ma}
\affiliation{Department of Engineering Physics, Tsinghua University, Beijing 100084, China}
\affiliation{Institute of High Energy Physics, Chinese Academy of Science, Beijing 100049, China}
\author{Jianfei Hua}
\email[]{jfhua@tsinghua.edu.cn}
\affiliation{Department of Engineering Physics, Tsinghua University, Beijing 100084, China}
\author{Dexiang Liu}
\affiliation{Department of Engineering Physics, Tsinghua University, Beijing 100084, China}
\author{Yunxiao He}
\affiliation{Department of Engineering Physics, Tsinghua University, Beijing 100084, China}
\author{Tianliang Zhang}
\affiliation{Department of Engineering Physics, Tsinghua University, Beijing 100084, China}
\author{Jiucheng Chen}
\affiliation{Department of Engineering Physics, Tsinghua University, Beijing 100084, China}
\author{Fan Yang}
\affiliation{Department of Engineering Physics, Tsinghua University, Beijing 100084, China}
\author{Xiaonan Ning}
\affiliation{Department of Engineering Physics, Tsinghua University, Beijing 100084, China}
\author{Hongze Zhang}
\affiliation{Department of Engineering Physics, Tsinghua University, Beijing 100084, China}
\author{Yingchao Du}
\affiliation{Department of Engineering Physics, Tsinghua University, Beijing 100084, China}
\author{Wei Lu}
\email[]{weilu@tsinghua.edu.cn}
\affiliation{Department of Engineering Physics, Tsinghua University, Beijing 100084, China}
\affiliation{Beijing Academy of Quantum Information Sciences, Beijing 100193, China}


\begin{abstract}
  Polarized X-ray source is an important probe for many fields such as fluorescence imaging, magnetic microscopy, and nuclear physics research. All-optical inverse Compton scattering source (AOCS) based on laser wakefield accelerator (LWFA) has drawn great attention in recent years due to its compact scale and high performance, especially its potential to generate polarized X-rays. Here, polarization-tunable X-rays are generated by a plasma-mirror-based AOCS scheme. The linearly and circularly polarized AOCS pulses are achieved with the mean photon energy of 60($\pm$5)/64($\pm$3) keV and the single-shot photon yield of $\sim$1.1/1.3$\times10^7$. A Compton polarimeter is designed to diagnose the photon polarization states, demonstrating AOCS’s polarization-tunable property, and indicating the average polarization degree of the linearly polarized AOCS is 75($\pm$3)\%.
\end{abstract}

\maketitle


\section{INTRODUCTION}
\label{Sec:1}

Polarized X-ray source has important applications in widespread fields, particularly in fluorescence imaging, magnetic microscopy, and nuclear physics research. For example, the utilization of a linearly polarized X-ray source in fluorescence imaging can significantly improve the image quality since the Compton scattering photon background can be greatly suppressed in the X-ray polarization direction \cite{RN114,RN113}. Left- and right-circularly polarized X-rays appear relatively different transmission in magnetic materials (magnetic circular dichroism), which is just the basic theory of magnetic microscopy \cite{RN115,RN181}, and this property is also especially advantageous in nuclear physics to study magnetic materials \cite{RN180}.

Polarized X-ray sources can be usually generated by synchrotron radiation, X-ray FEL, bremsstrahlung radiation, and inverse Compton scattering (ICS). Among these mechanisms, synchrotron radiation \cite{RN118,RN176} and X-ray FEL \cite{RN185,RN184} can produce ultra-bright monoenergetic $\sim$100\%-polarized X-rays, and their polarization states can be tuned by inserting different kinds of undulators, however, they are large-scale facilities and can hardly access to high photon energy (typically of 1-100 keV). Bremsstrahlung radiation can produce keV- to MeV-level linearly polarized X-rays at large observation angle when a thin target is utilized \cite{RN177,RN117}, and can also produce circularly polarized X-rays using longitudinally-polarized incident electrons \cite{RN178}, however, these photons are typically partially polarized and the polarization state adjustment is complicated. ICS can produce keV-MeV quasi-monoenergetic X-rays with high polarization degrees and their polarization states can be easily adjusted by changing the polarization state of the scattering laser pulse, showing distinct advantages in practical application \cite{RN121,RN146,RN119}.

\begin{figure}[ht]
  \includegraphics[width=8.5cm]{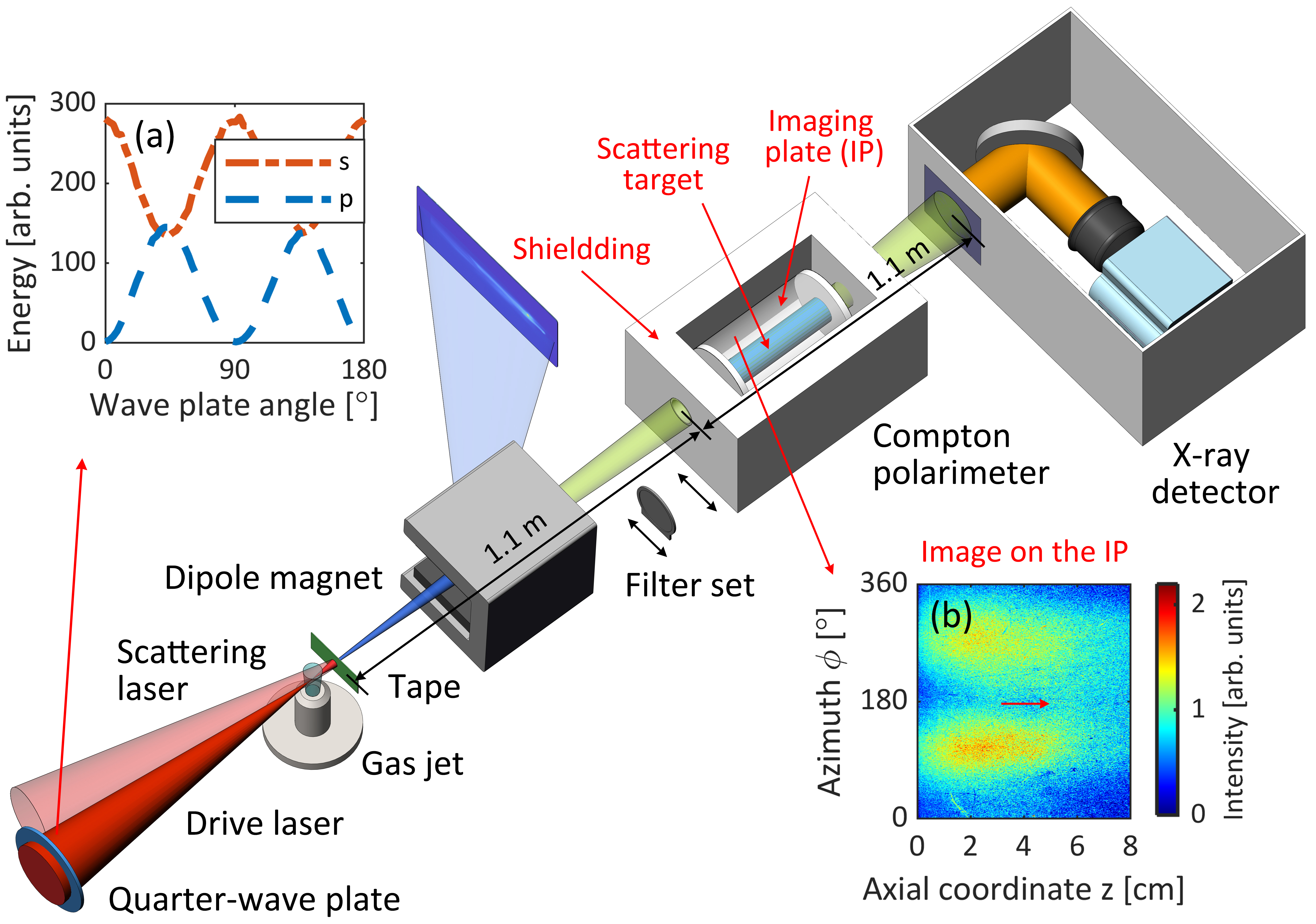}
  \caption{Schematic layout of the experiment, where the filter set and the Compton polarimeter are movable. (a) Laser energy of the s- and p-polarized components at different wave plate angles, which shows that the drive laser is linearly polarized at $0^{\circ}$, circularly polarized at $45^{\circ}$, and elliptically polarized between $0^{\circ}$ to $45^{\circ}$. (b) The measured image on the imaging plate in the Compton polarimeter, where the red arrow shows the propagation direction of the incident AOCS photons.}
  \label{Fig:1}
\end{figure}

\begin{figure*}[ht]
  \includegraphics[width=15cm]{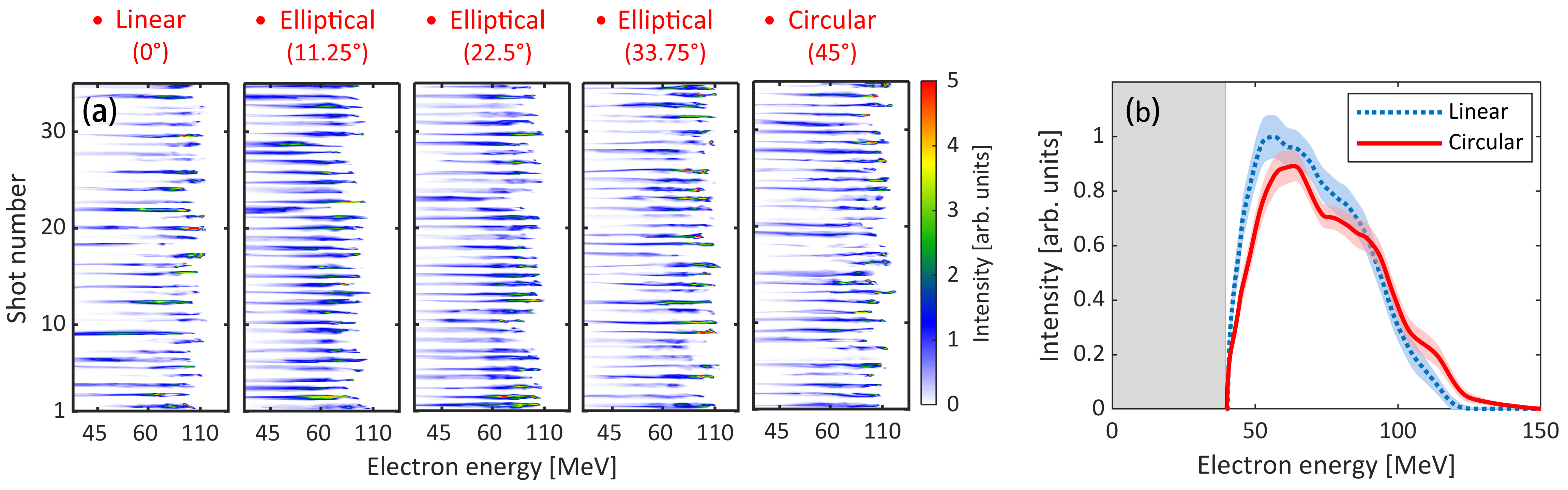}
  \caption{(a) The electron beam spectra for continuous 35 shots driven by different-polarization laser pulses, where the wave plate angle for each polarization state is labelled on the top. (b) 50-shots accumulated spectra of the electron beams driven by linearly and circularly polarized laser pulses, where the shaded areas around the curves show their FWHM errors and the detection threshold is highlighted by a gray rectangle.}
  \label{Fig:2}
\end{figure*}

Laser wakefield accelerator (LWFA)\cite{RN47,RN14,RN75} can generate MeV-GeV electron beams on a tabletop due to its ultra-high acceleration gradient about three orders of magnitude higher than that of conventional accelerators. ICS based on LWFA, termed all-optical inverse Compton scattering source (AOCS) \cite{RN27,RN28,RN77,RN33,RN34}, has drawn great attention due to its high performance (polarization-tunable, quasi-monoenergetic, energy-tunable) and benchtop scale. To date, several experiments about the optimization and imaging application of AOCS are reported, while its polarization properties have not been experimentally demonstrated.

In this paper, the generation of polarization-tunable X-rays is demonstrated using a plasma-mirror-based AOCS scheme \cite{RN33,RN34}, and the polarization states of linearly and circularly polarized X-rays are diagnosed to verify the polarization-tunable property. In the plasma-mirror-based AOCS scheme, the drive laser of LWFA is reflected by a plasma mirror to scatter with the energetic LWFA electron beam, thus both the temporal and spatial synchronization can be easily realized, especially simplifying the source system. The generated AOCS X-ray photons within $1/\gamma$ divergence duplicate the polarization state of the scattering laser \cite{RN146}, and the on-axis X-ray photons have the highest polarization degree (up to $\sim$100\%), thus the polarization state of AOCS photons can be adjusted by changing the scattering laser polarization. A dedicated Compton polarimeter \cite{RN129,RN127} is designed to measure the polarization state of the AOCS X-rays, showing good accordance with the simulation results.

\section{POLARIZED AOCS X-RAYS GENERATION}
\label{Sec:2}

The experiment is performed on the 10 TW laser system in Tsinghua University \cite{RN69}, and the setup is sketched in FIG. \ref{Fig:1}. Intense ultra-short laser pulses (500 mJ, 40 fs) are focused to an FWHM spot size of 9 $\mu \rm m$ (65\% energy enclosed) near the leading edge of the gas jet (2 mm) by an f/18 off-axis parabolic mirror. A quarter-wave plate (diameter of 50 mm) is inserted in the laser path upstream of the OAP to adjust the polarization state of the laser pulses by changing the wave plate angle (angle between the optical-axis of the quarter-wave plate and the polarization direction of incident laser), as shown in FIG. \ref{Fig:1}(a). When a laser pulse passes through the working gas (99.5\% He + 0.5\% $\rm {N_2}$) ejected from the gas jet, plasma with the density of $\sim$1$\times10^{19} \rm {cm^{-3}}$ is generated by laser ionization, and the plasma wakefield is then formed behind the laser pulse. A mass of electrons is injected into the acceleration phase of the wakefield and then accelerated continuously to high energy. To measure the spectrum of the electron beam, an electron spectrometer based on a dipole magnet (magnetic field of 1 T) is inserted in the beam path. Electron beam spectra for continuous 35 shots driven by laser pulses with different polarization states are shown in FIG. \ref{Fig:2}(a), indicating that stable electron beams can be accelerated using all these polarized laser pulses. For linearly and circularly polarized drive lasers, the generated electron beams have divergences of $\sim$17/13 mrad and $\sim$18/16 mrad (horizontal/vertical), and charges of 52 pC (36\% RMS-jitter) and 74 pC (36\% RMS-jitter), respectively. The corresponding 50-shots accumulated electron beam spectra are shown in FIG. \ref{Fig:2}(b), with average energies of 70.3 MeV (72\% energy spread) and 74 MeV (70\% energy spread) for linearly and circularly polarized cases respectively.

\begin{figure}[ht]
  \includegraphics[width=8.4cm]{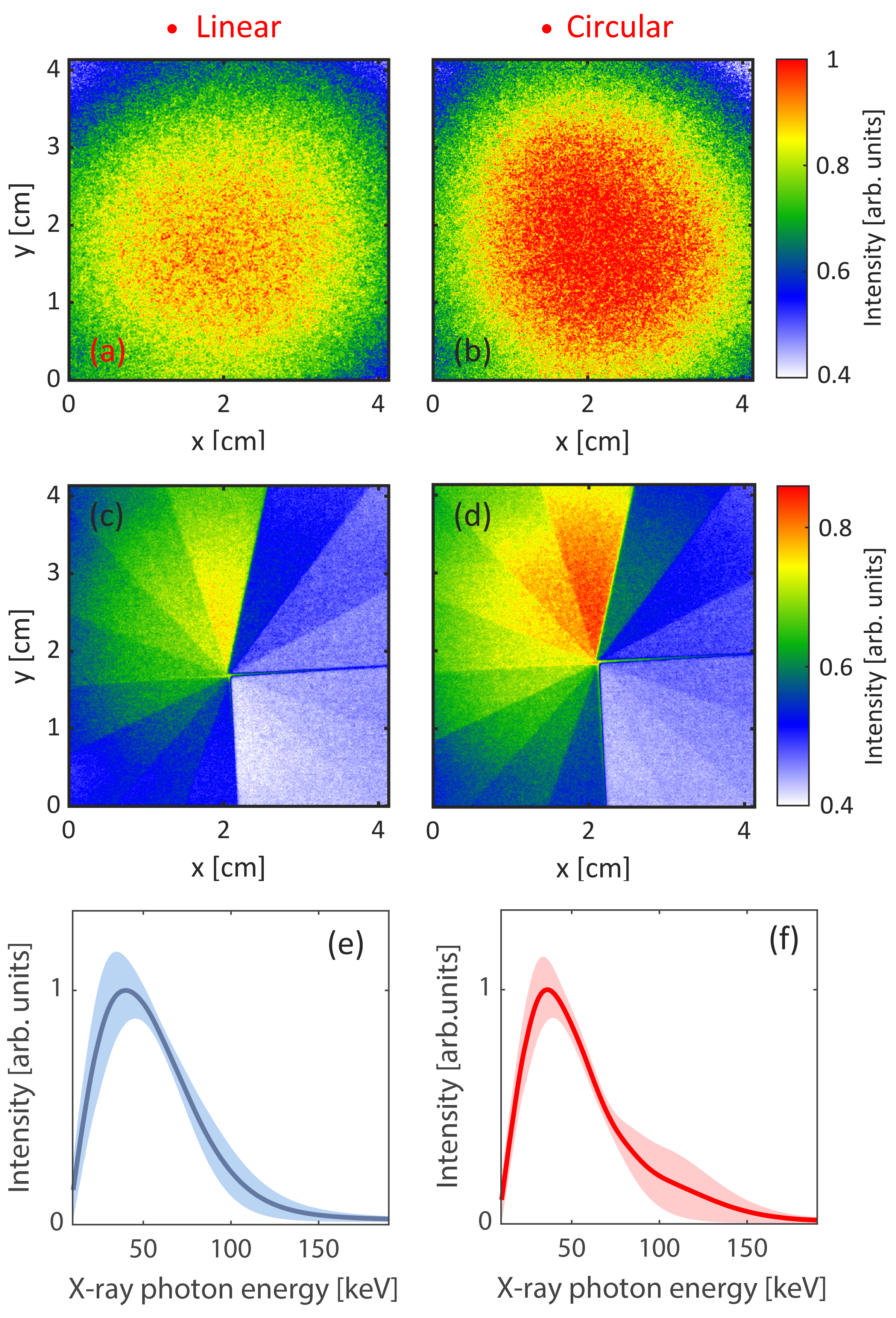}
  \caption{(a) and (b) are the profiles of linearly- and circularly-polarized AOCS X-rays detected by the X-ray detector, while (c) and (d) are those attenuated by the filter set. (e) and (f) are the reconstructed spectra of linearly- and circularly-polarized AOCS X-rays, where the shaded areas around the two curves are the reconstruction errors.}
  \label{Fig:3}
\end{figure}

A tape system equipped with 100-$\mu \rm m$-thick polyethylene terephthalate (PET) film is placed right behind the gas jet, thus a plasma mirror is formed when the drive laser hits the PET film. The residual drive laser then collides with the LWFA electron beam after being reflected by the plasma mirror. Since the laser pulse will conserve its polarization state during both the LWFA process and the $\sim$180-degree reflection, X-rays with the same polarization state as the scattering laser will be generated by the ICS process. Therefore, polarization-tunable X-rays are achieved by straightforwardly adjusting the polarization state of the drive laser pulses.

The generated AOCS X-rays are detected by a calibrated CsI(Tl) scintillation screen (Hamamatsu J8734) imaged onto an EMCCD (ANDOR DU888E), as shown in FIG. \ref{Fig:3}. The projection profiles of linearly and circularly polarized X-rays are illustrated in FIG. \ref{Fig:3}(a) and (b), showing that their divergences are $\sim$37/33 mrad and $\sim$29/28 mrad (horizontal/vertical) respectively. To measure the X-ray spectrum, a filter set composed of different materials with different thicknesses (7 aluminum plates with the thickness from 0.35 mm to 8 mm and 7 copper plates with the thickness from 0.35 mm to 9 mm) is inserted on the X-ray path \cite{RN56,RN179}. For certain materials, the X-ray attenuation coefficients vary according to photon energy, thus the spectral information will be recorded in the attenuated X-ray profile by the filter set, as shown in FIG. \ref{Fig:3}(c) and (d). The X-ray spectra are then reconstructed by the Expectation-Maximization (EM) algorithm \cite{RN125} using the data in \ref{Fig:3}(c) and (d), as shown in FIG. \ref{Fig:3}(e) and (f), indicating that the average photon energies are 60($\pm$5) keV and 64($\pm$3) keV for linearly and circularly polarized X-rays respectively. Using the X-ray profiles and spectra above, and according to the calibrated efficiency of the X-ray detector, the photon yields are then estimated to be $\sim$1.1$\times10^7$ (linear polarization) and $\sim$1.3$\times10^7$ (circular polarization) per shot.

\section{POLARIZATION STATE MEASUREMENT}
\label{Sec:3}

To diagnose the polarization states of AOCS pulses, a Compton polarimeter is designed according to our X-ray parameters \cite{RN129,RN127}. Compton scattering process happens when X-rays interact with matters, and the differential cross-section ${d \sigma}/{d \Omega}$ can be described as

\begin{equation}
  \begin{aligned}
  \frac{d \sigma}{d \Omega}&=\frac{r_{e}^{2}}{2} \frac{E^{2}}{E_{0}^{2}}\\&\left(\frac{E}{E_{0}}+\frac{E_{0}}{E}-2 \sin ^{2} \theta \frac{\cos ^{2} \phi+a \sin ^{2} \phi}{a+1}\right)
  \end{aligned}
  \label{Eq:1}
\end{equation}

\noindent where $r_e$ is the classical electron radius, $E_0$, and $E$ are the photon energy of the incident X-rays and the secondary Compton X-rays, $\theta$ is the forward scattering angle, $\phi$ is the azimuthal scattering angle, and $a$ is related to the polarization state of incident X-rays ($a=0$ or $a \rightarrow \infty$ for linear polarization, $a=1$ for circular polarization). Eq. \eqref{Eq:1} shows that the azimuthal distribution of the secondary Compton photons at $\theta = \pi /2$ is $\cos^2\phi$ (or $\sin^2\phi$) related for the linearly polarized incident X-rays, while uniform for the circularly polarized incident X-rays, thus this characteristic can be used to identify different X-ray polarization states. Besides, this characteristic can also help to calculate the polarization degree of linearly polarized X-rays by comparing the amplitude of its $\cos^2 \phi$-related distribution with that of 100\%-linearly-polarized X-rays.

\begin{figure}[ht]
  \includegraphics[width=8.4cm]{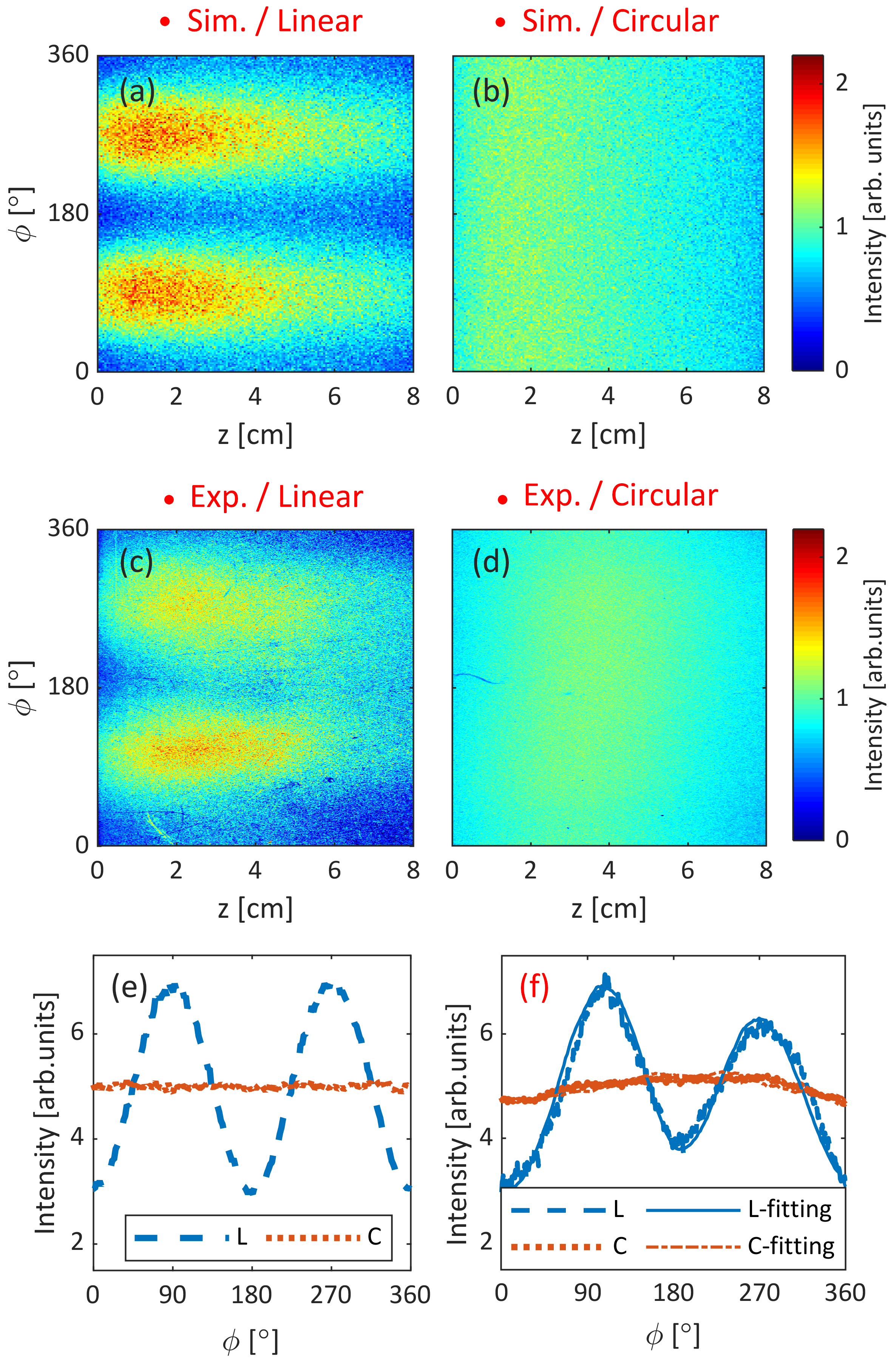}
  \caption{(a) and (b) are the simulation results of the Compton polarimeter for 100\% linearly and circularly polarized incident X-rays, and the corresponding one-dimensional distributions are plotted in (e). (c) and (d) are the experimentally measured results of the Compton polarimeter for linearly- and circularly-polarized incident AOCS photons, and the corresponding one-dimensional distributions along with the simulated fitting curves are plotted in (f).}
  \label{Fig:4}
\end{figure}

The structure of the Compton polarimeter is shown in FIG. \ref{Fig:1}. To improve the secondary Compton photon flux and reduce self-absorption, a polyethylene cylinder is chosen to be the scattering target due to its high Compton scattering cross-section (0.186 $\rm {g/cm^2}$) and low density (0.96 $\rm {g/cm^3}$). Considering the X-ray photon energy (60$\sim$64 keV), divergence (28$\sim$37 mrad), and drift distance (1.1 m), the polyethylene target is designed to have a 2-cm diameter (corresponding to 18.2 mrad acceptance divergence for the incident X-rays) and 10-cm length (attenuated to 15\% for 60 keV X-ray photon). When polarized X-ray photons pass through the scattering target along its axis, Compton photons will be generated and their azimuthal distributions are detected by an imaging plate (IP) rolled around the scattering target. It should be noted that there are two 1-cm-thickness mounts at both ends of the polyethylene target to hold the IP, thus the effective axial detection length is 8 cm. A shielding system composed of an inner layer of 5-cm-thickness polyethylene and an outer layer of 10-cm-thickness lead encloses the scattering target and IP, with a 2-cm-diameter hole in the front end to collimate the X-rays. To preliminarily test this polarimeter, a Monte Carlo simulation is performed by FLUKA \cite{RN126}. In the simulation, 100\% linearly and circularly polarized X-rays with the experimentally measured parameters (photon energy and divergence) are adopted, and the polarimeter setup is the same as introduced above. The simulated two-dimensional and one-dimensional azimuthal distributions of the Compton photons are illustrated in FIG. \ref{Fig:4}(a), (b), and (e), where the $\cos^2 \phi$-related distribution for the linearly polarized case and uniform distribution for the circularly polarized case are both evident, showing this polarimeter is indeed feasible for X-ray polarization diagnosis.

In the experiment, continuous 600 shots linearly and circularly polarized AOCS pulses are performed in each measurement, and the corresponding two-dimensional azimuthal distributions of the secondary Compton photons on the IP are shown in FIG. \ref{Fig:4}(c) and (d) respectively. The one-dimensional distributions of the linear and circular polarization measurements are plotted in FIG. \ref{Fig:4}(f), showing similar characteristics with the simulation results in FIG. \ref{Fig:4}(a) and (b). In the Compton polarimeter, the amplitude of the $\cos^2 \phi$-related distribution for the linearly polarized X-rays depends on the polarization degree: smaller amplitude for lower polarization degree. Besides, for both linearly and circularly polarized X-rays, once they are not aligned well with the scattering target, more secondary Compton photons will be generated in the direction that the X-rays are tilted, thus the amplitudes of the $\cos^2 \phi$-related distribution for linearly polarized X-rays will become uneven and the secondary photon distribution of circularly polarized X-rays will become non-uniform. Considering the above two factors, another two FLUKA simulations are performed to fit the experimental curves by changing the polarization degree of the X-rays (only for linear polarization) and adjusting the offset between the X-ray path with the scattering target. The best-fitting curves are plotted in FIG. \ref{Fig:4}(f), indicating that the linearly polarized X-rays have $75(\pm 3)\%$ polarization degree and deviate from the axis of the scattering target by $\sim$8.5 mm/5 mm (horizontal/vertical), and the circularly polarized X-rays deviate by $\sim$3.5 mm (distance between their axes).

\begin{figure*}[ht]
  \includegraphics[width=15cm]{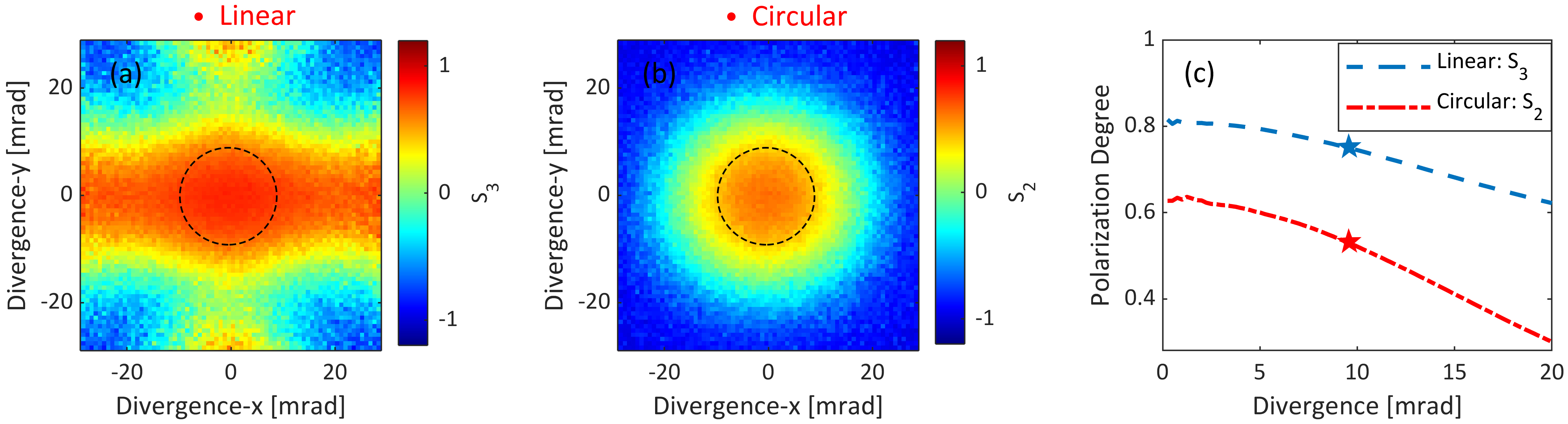}
  \caption{(a) and (b) are the two-dimensional polarization degree distributions of the linearly ($S_3$) and circularly ($S_2$) polarized AOCS X-rays, where the black dashed circles indicate the detection area of the Compton polarimeter. The average polarization degrees (the average of $S_3$ for linear polarization and $S_2$ for circular polarization) within different divergences are plotted in (c), where the pentacles indicate the measured value of our Compton polarimeter.}
  \label{Fig:5}
\end{figure*}

To further check the polarization degree measured above, another two Monte Carlo simulations about the AOCS generation are performed by CAIN code \cite{RN57} according to the experiment parameters. The scattering lasers are separately set to have 100\% linear and circular polarization degree, with $a_0=1.3$ (linear polarization) or $1.3 / \sqrt{2}$ (circular polarization) and the wavelength of 900 nm (considering the evolution of the drive laser during the acceleration process and reflection process) \cite{RN71,RN179}. The polarization degree of linearly and circularly polarized X-rays are described by Stokes parameter $S_3$ and $S_2$ respectively \cite{RN146}. The simulated two-dimensional distributions of $S_3$ and $S_2$ are shown in FIG. \ref{Fig:5}(a) and (b), where the detection area in the experiment (18.2 mrad) is highlighted by black dashed circles. The average polarization degrees (corresponding to the average of $S_3$ and $S_2$) within different divergences are plotted in FIG. \ref{Fig:5}(c), showing that a higher average polarization degree can be obtained within smaller divergence. At the acceptance divergence of our Compton polarimeter, the average polarization degrees for linearly/circularly-polarized AOCS photons are 76\%/54\% respectively, where the linear polarization degree agrees well with the experimentally measured value ($75(\pm 3)\%$).

\section{CONCLUSIONS}
\label{Sec:4}

In this paper, the generation of $\sim$60-keV polarization-tunable X-rays using the plasma-mirror-based AOCS scheme is demonstrated, and the polarization states of linearly and circularly polarized X-rays are diagnosed by a Compton polarimeter to verify the polarization-tunable property of this source. The polarization degree of the linearly polarized X-rays is measured to be $\sim$75\%, consistent with the Monte Carlo simulation. The polarization state of our AOCS source can be easily adjusted by tuning the drive laser polarization using a quarter-wave plate, which is especially efficient and convenient for practical application. Polarization of AOCS X-rays is closely related to the emittance of electron beams \cite{RN146}, thus the polarization degree of this source can be further improved by optimizing the LWFA electron beam. This source can also generate $\sim$MeV polarization-tunable X-rays through upgrading the laser system, which especially contributes to high-energy applications such as nuclear physics research. To summarize, AOCS is a potential table-top X-ray source for many applications, and our demonstration of its polarization-tunable property shows extreme advantages for wider scientific areas.

\begin{acknowledgments}
The authors thank Dr. Yuchi Wu for the support of X-ray detection. This work is supported by the National Natural Science Foundation of China Grants, No. 11991071, No. 11875175, No. 11991073, and Tsinghua University Initiative Scientific Research Program.
\end{acknowledgments}

\bibliography{myref}

\end{document}